\documentclass[prd,aps,twocolumn,showpacs,preprintnumbers,amsmath,amssymb]{revtex4-2}
\usepackage[utf8]{inputenc}
\usepackage{graphicx}
\usepackage[breaklinks, colorlinks=true]{hyperref}
\usepackage{bm}
\usepackage{physics}
\usepackage{siunitx}
\usepackage{mathtools}
\usepackage{here}
\usepackage{amsmath}

\newcommand{\ri}{\mathrm{i}}

\newcommand{\Lag}{\mathcal{L}}

\begin{document}

\title{Pseudogauge ambiguity in the distributions of energy density,\\
pressure, and shear force inside the nucleon}

\author{Kenji Fukushima}
\email{fuku@nt.phys.s.u-tokyo.ac.jp}
\affiliation{Department of Physics, The University of Tokyo, 
  7-3-1 Hongo, Bunkyo-ku, Tokyo 113-0033, Japan}

\author{Tomoya Uji}
\email{uji@nt.phys.s.u-tokyo.ac.jp}
\affiliation{Department of Physics, The University of Tokyo, 
  7-3-1 Hongo, Bunkyo-ku, Tokyo 113-0033, Japan}

\begin{abstract}
We study the spatial distributions of pressure, energy density, and shear forces inside the nucleon within the two-flavor Skyrme model including vector mesons.  This framework has the advantage that nucleon configurations can be stabilized without the Skyrme term.  In contrast to the model without vector mesons, however, we realize that the energy–momentum tensor (EMT) becomes pseudo-gauge dependent.  We explicitly demonstrate that all these distributions differ between the canonical and Belinfante forms of the EMTs.  We identify the pseudo-gauge ambiguity as originating from nonvanishing surface terms associated with spin currents generated by the vector-meson field strength tensors.  Furthermore, we show that the pressure and shear-force distributions in the canonical EMT develop singularities at the nucleon center, whereas the corresponding Belinfante distributions remain finite.  Finally, we discuss the implications of pseudo-gauge dependence for extracting the confining force and for constructing the equation of state inside the nucleon.
\end{abstract}

\maketitle

\section{Introduction}
Understanding the internal structure of the nucleon, including its mass, spin, and distributions of confining forces, has become a central subject in QCD physics~\cite{Ji:1994av, Ji:1996ek, Polyakov:2002yz, Aidala:2012mv, Polyakov:2018zvc}. These properties are encoded in the matrix elements of the energy-momentum tensor (EMT), which can be parametrized by a set of EMT form factors. Among the most promising experimental tools to access these form factors are hard exclusive processes, such as deeply virtual Compton scattering (DVCS)~\cite{Muller:1994ses, Radyushkin:1996nd, Ji:1996nm, Collins:1996fb, Vanderhaeghen:1998uc, Ji:1998pc, Belitsky:2001ns, Goloskokov:2005sd, Kumericki:2007sa, Kroll:2014tma}, which probe generalized parton distributions (GPDs)~\cite{Burkardt:2002hr, Diehl:2003ny, Belitsky:2005qn}.
The upcoming Electron-Ion Collider (EIC)~\cite{Accardi:2012qut, Abir:2023fpo} is expected to advance the experimental determination of GPDs significantly.
In particular, the $D$-form factor, one of the EMT form factors, has attracted attention as it encodes information about the internal mechanical forces, and the $D$-term, forward value of the $D$-form factor, is often referred to as ``the last unknown global property'' of the nucleon~\cite{Polyakov:1999gs, Goeke:2007fp, Cebulla:2007ei, Shanahan:2018nnv, Lorce:2018egm, Burkert:2018bqq, Chakrabarti:2020kdc, Burkert:2021ith, Ji:2021mfb, Fujita:2022jus, Burkert:2023wzr, GarciaMartin-Caro:2023klo, GarciaMartin-Caro:2023toa, Lorce:2025oot, Polyakov:2018zvc, Guo:2025jiz, Goharipour:2025lep, Ji:2025gsq, Sugimoto:2025btn, Tanaka:2025pny, Ji:2025qax, Stegeman:2025sca}.

The physical interpretation of the $D$-form factor has been formulated in terms of the EMT in the rest frame of the nucleon, leading to a mechanical picture involving pressure and shear force distributions.
From both phenomenological analyses and theoretical arguments, the radial pressure is positive everywhere in the nucleon, and the shear force, rather than the pressure itself, is responsible for the inward confining force that balances the outward pressure gradient~\cite{Burkert:2018bqq, Lorce:2018egm}.  This mechanical equilibrium inside the nucleon has no doubt, for it is the consequence of the hydrostatic equilibrium equation derived from the conservation law of the EMT\@.  However, the specific values of pressure and shear force are not unique because the definition of EMT can be changed while satisfying the conservation law~\cite{Ji:2025gsq}.  In other words, there is pseudo-gauge ambiguity about the EMT definition.

The pseudo-gauge transformation of the EMT has been discussed extensively; see Refs.~\cite {Hehl:1976vr, Leader:2013jra} for reviews and Refs.~\cite{Fukushima:2020ucl,Li:2020eon,Drogosz:2024rbd,Becattini:2025twu} for discussions in the context of spin hydrodynamics.  The readers who are interested in spin hydrodynamics can consult reviews~\cite{Florkowski:2018fap,Bhadury:2025dzh}.
There are two commonly used EMTs: one is the canonical EMT, which is straightforwardly obtained by Noether's theorem, and the other is the Belinfante improved EMT, which is refined by the pseudo-gauge transformation from the canonical one.  In the gauge theory, the Belinfante EMT seems a natural choice because it is symmetric and gauge invariant, whereas the canonical EMT is asymmetric and not gauge invariant.  Therefore, discussing the internal properties of hadrons, transverse momentum distributions (TMDs) and GPDs corresponding to the Belinfante EMT are usually used~\cite{Ji:1996ek, Polyakov:2018zvc}.  However, by changing the path of the Wilson line appearing in TMDs and GPDs, those distributions corresponding to the canonical EMT may become gauge invariant and physical observables~\cite{Lorce:2015lna}.

The canonical EMT takes a crucial role in spin physics.  The famous Jaffe-Manohar spin decomposition~\cite{Jaffe:1989jz} is based on the canonical and gauge-dependent angular momentum operators, and they have been improved to an explicitly gauge-independent form by introducing nonlocal operators~\cite{Chen:2008ag, Wakamatsu:2010qj, Hatta:2011zs, Leader:2015vwa}.  Moreover, in the context of spin hydrodynamics, the anti-symmetric part of the EMT, which is absent in the Belinfante EMT, is important to describe the spin degrees of freedom~\cite{Becattini:2011ev, Li:2020eon,Drogosz:2024rbd,Becattini:2025twu, Becattini:2012pp, Hattori:2019lfp, Fukushima:2020qta, Speranza:2020ilk, Fukushima:2020ucl, Fang:2025aig}.  Apart from these representative choices, depending on the context, an alternative pseudo-gauge may be more useful, e.g., the choice by de~Groot, van~Leeuwen, and van~Weert called the GLW pseudo-gauge~\cite{Florkowski:2018fap}.

In this work, we extensively study the aforementioned two types of EMTs inside the nucleon using the two-flavor Skyrme model with vector mesons~\cite{Adkins:1983nw, Igarashi:1985et, Meissner:1986ka, Meissner:1986js}, and we show the energy density, the pressure, and the shear force distributions extracted from both EMTs.
Unlike the ordinary Skyrme model, this model contains the vector mesons in the form of gauge fields, so that a difference can be manifested between the canonical EMT and Belinfante EMT\@.
Usually, in the presence of dynamical gauge fields, the canonical EMT is unfavorable from the perspective of gauge invariance.
However, the vector meson field is not a true gauge field. 
Thus, we cannot immediately conclude that the Belinfante EMT in this model is more physical.

Our main finding from this model analysis is that the local values of energy density $\varepsilon(r)$, pressure $p(r)$, and shear force $s(r)$ can be changed via the pseudo-gauge transformation.
Thus, the relation between them, such as $p=p(\varepsilon)$, which is referred to as the equation of state (EoS)~\cite{Rajan:2018zzy}, can be modified.
However, the macroscopic stability condition is unchanged, that is, both forms of pressure satisfy the von~Laue condition~\cite{Laue:1911lrk}, which is the necessary condition for the nucleon stability derived from the virial theorem~\cite{Polyakov:2018zvc}.
The pseudo-gauge difference takes the form of total derivative terms.  Consequently, conserved charges after spatial integration have no dependence on pseudo-gauge, provided that the fields fall off sufficiently rapidly at spatial infinity. In other words, we limit our consideration to sufficiently smooth pseudo-gauge transformations that do not generate topological defects.
We note, however, that physical integrals with some weight functions can be sensitive to the pseudo-gauge choice such as the surface tension energy.

Our results support the recent statement made in Ref.~\cite{Ji:2025gsq} that local distributions inferred from the EMT form factors do not seem to have any overarching mechanical significance.
Although it may be possible to constrain some pseudo-gauge choices from gauge invariance, there seems to be no physical guiding principle to find the most advantageous form of EMT in a unique way.

This paper is organized as follows.  In Sec.~\ref{sec:theoretical}, we make a brief review of the theoretical framework including the energy-momentum tensor and the pseudo-gauge transformation.  We also write down the equations of motion for fields of the pion, the $\rho$ meson, and the $\omega$ meson to solve the soliton solution in the Skyrme model.  Our results are presented in Sec.~\ref{sec:results}, where the difference between the canonical and the Belinfante EMTs is closely discussed.  In Sec.~\ref{sec:discussions}, we then apply the resulting distributions of energy density, pressure, and shear force for physical discussions on the confining force as well as the EoS of dense matter.  Section~\ref{sec:conclusions} is devoted to the conclusions.

\section{Theoretical framework}\label{sec:theoretical}
We summarize the formulae relevant to the present discussion, namely, the explicit forms of different EMTs and the definition of the Skyrme model with vector mesons.

\subsection{Energy-momentum tensor}
\begin{figure}
    \centering
    \includegraphics[width=0.5\linewidth]{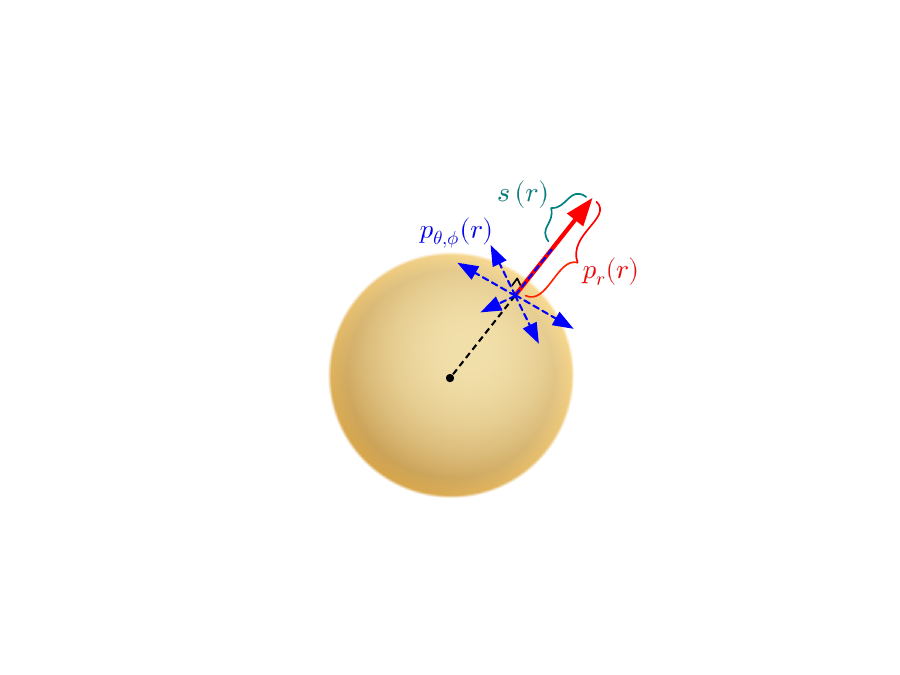}
    \caption{Schematic illustration for $p_r(r)$ and $p_{\theta,\phi}(r)$.}
    \label{fig:pr_ptheta}
\end{figure}

We decompose the EMT into the energy density, the pressure, and the shear force inside a nucleon as usual~\cite{Polyakov:2018zvc}:
\begin{equation}
\begin{split}\label{eq:EMT}
    T^{00} &= \varepsilon(r)\,, \\
    T^{ij} &= \delta^{ij}p(r) + \left( \hat{r}^i \hat{r}^j - \frac{1}{3}\delta^{ij}\right)s(r)\,.
\end{split}
\end{equation}
We define a unit vector as $\hat{\bm r}={\bm r}/|{\bm r}|$.
Then, we regard a nucleon as a spherically symmetric system, so that we can diagonalize $T^{ij}$ and define the radial pressure, $p_r$, and the tangential pressure, $p_{\theta,\phi}$ as
\begin{equation}
    p_r(r) = p(r) + \frac{2}{3}s(r)\,,\qquad p_{\theta,\phi}(r) = p(r) - \frac{1}{3}s(r)\,.
\end{equation}
The graphical interpretation of this pressure decomposition is given in the schematic illustration in Fig.~\ref{fig:pr_ptheta}.

From the conservation law, $\partial_\mu T^{\mu\nu} = 0$, we get a mechanical equilibrium condition~\cite{Polyakov:2018zvc, Lorce:2018egm}, which corresponds to the hydrostatic equation to describe the NS structures:
\begin{equation}\label{eq:hydroequilibrium}
    p_r^\prime(r) = -\frac{2}{r}s(r)\,.
\end{equation}
This condition indicates that the shear force plays the role of the inward confining force that balances the outward pressure gradient.
This balance equation takes the same form as the NS hydrostatic equation with the gravitational force replaced with the confining force.

We can modify the EMT not violating the conservation law, i.e., $\partial_\mu T^{\mu\nu} = 0$, as
\begin{equation}\label{eq:pseudogauge}
    \tilde{T}^{\mu\nu} = T^{\mu\nu} + \partial_\lambda\mathcal{K}^{\lambda\mu\nu}\,.
\end{equation}
Here, we can easily confirm $\partial_\mu \tilde{T}^{\mu\nu} = 0$ as long as an arbitrary tensor $\mathcal{K}^{\lambda\mu\nu}$ is antisymmetric with respect to $\lambda$ and $\mu$.
The redefinition of the EMT in Eq.~\eqref{eq:pseudogauge} is called the pseudo-gauge transformation.
Although $T^{\mu\nu}$ is not uniquely defined, the most familiar form is the canonical EMT derived from Noether's theorem.
The explicit form of the canonical EMT is 
\begin{equation}
    T_\text{can}^{\mu\nu} = \sum_{\phi}\frac{\partial\Lag}{\partial(\partial_\mu\phi)}\partial^\nu\phi - g^{\mu\nu}\Lag\,,
    \label{eq:Tcan}
\end{equation}
where $\Lag$ is a Lagrangian density in terms of general fields denoted by $\phi$.
The canonical EMT is not generally symmetric with respect to $\mu$ and $\nu$.
Alternatively, there is a definition of the EMT based on metric differentiation:
\begin{equation}
    T^{\mu\nu} = -2\pdv{\Lag}{g_{\mu\nu}} - g^{\mu\nu}\Lag\,,
    \label{eq:TBel}
\end{equation}
which is symmetric by definition.
This EMT often coincides with the Belinfante form.
These different forms of the EMTs are related via the following pseudo-gauge transformation: 
\begin{equation}
    T_\text{Bel}^{\mu\nu} = T_\text{can}^{\mu\nu} + \frac{1}{2}\partial_\lambda(S^{\lambda\mu\nu} - S^{\mu\lambda\nu} + S^{\nu\mu\lambda})\,,
\end{equation}
where $S^{\lambda\mu\nu}$ is the spin current operator.

\subsection{Skyrme model with vector mesons}
We adopt the two-flavor Skyrme model with vector mesons~\cite{Adkins:1983nw, Igarashi:1985et, Meissner:1986ka, Meissner:1986js}.
The Lagrangian density is
\begin{align}\label{eq:Lagrangian}
    \Lag &= \frac{1}{4}f_\pi^2\tr(\partial_\mu U \partial^\mu U^\dagger) - \frac{1}{2}f_\pi^2\tr(D_\mu\xi\cdot\xi^\dagger+D_\mu\xi^\dagger\cdot\xi)^2 \notag \\
    &\qquad - \frac{1}{2g^2}\tr(F_{\mu\nu}F^{\mu\nu}) + \frac{3}{2} g\omega_\mu B^\mu \notag \\
    &\qquad + \frac{1}{2}m_\pi^2f_\pi^2\tr(U + U^\dagger - 2)
    \,,
\end{align}
where $U$ is the nonlinear representation of the pions satisfying $U = \xi^2$, and $\omega_\mu$ is the isosinglet vector meson.
Here, $B^\mu$ is a baryon current:
\begin{equation}
    B^\mu = \frac{1}{24\pi^2}\varepsilon^{\mu\nu\alpha\beta}\tr(U^\dagger\partial_\nu UU^\dagger\partial_\alpha UU^\dagger\partial_\beta U)\,.
\end{equation}
In the Lagrangian density, $D_\mu$ and $F_{\mu\nu}$ are the covariant derivative and the field strength tensor with vector meson fields, given respectively as
\begin{align}
    D_\mu &= \partial_\mu - \ri V_\mu\,,\\
    F_{\mu\nu} &= \partial_\mu V_\nu - \partial_\nu V_\mu - \ri[V_\mu, V_\nu]\,.
\end{align}
In the above expressions, $V_\mu$ represents the vector meson matrix in flavor space defined by
\begin{equation}
     V_\mu = \frac{g}{2} (\bm{\tau}\cdot\bm{\rho}_\mu + \omega_\mu)
\end{equation}
with $\bm{\rho}_\mu$ the isotriplet vector mesons and ${\bm \tau}$ the $2\times 2$ Pauli matrices in flavor space.

To minimize the energy functional calculated from Eq.~\eqref{eq:Lagrangian}, we shall introduce Ans\"atze as follows.
We take the hedgehog Ansatz for the pions as
\begin{equation}
\label{eq:ansatz}
    \xi(\bm{r}) = \exp[\frac{1}{2}\ri\bm{\tau}\cdot\hat{\bm{r}}F(r)]\,,\qquad U(\bm{r}) = \exp[\ri\bm{\tau}\cdot\hat{\bm{r}}F(r)]\,.
\end{equation}
We choose the Wu-Yang-'t~Hooft-Polyakov Ansatz for $\rho$ mesons:
\begin{equation}\label{eq:rho}
    \rho^0(\bm{r}) = 0\,,\qquad
    \rho^{i, a}(\bm{r}) = \varepsilon^{ika}\hat{r}^k\frac{G(r)}{gr}\,.
\end{equation}
Since $\omega^\mu$ couples to the baryon density, $B^0$, at rest, we employ the following Ansatz:
\begin{equation}\label{eq:omega}
    \omega^\mu(\bm{r}) = \omega(r)\delta^{\mu 0}\,.
\end{equation}
From the conditions to minimize the energy, we get a set of equations of motion as follows~\cite{Meissner:1986js}:
\begin{align}
    F^{\prime\prime} &= - \frac{2}{r}F^\prime - \frac{3g}{4\pi^2f_\pi^2r^2}\omega^\prime\sin^2F \notag \\
    &\qquad + \frac{1}{r^2}[4(G + 1)\sin F - \sin2F] + m_\pi^2\sin F
    \,, \\
    G^{\prime\prime} &= \frac{1}{r^2}G(G + 1)(G + 2) + 2g^2f_\pi^2(G + 1 - \cos F)\,, \\
    \omega^{\prime\prime} &= -\frac{2}{r}\omega^\prime + 2g^2f_\pi^2\omega - \frac{3g}{4\pi^2r^2}F^\prime \sin^2 F\,.
\end{align}
To quantize the baryon number properly, the boundary condition for the pions should be
\begin{equation}
    F(0) = \pi\,,\qquad F(\infty) = 0\,.
\end{equation}
To avoid unphysical divergence in the energy, we impose the boundary conditions for the vector mesons as
\begin{align}
    &G(0) = -2\,, && \hspace*{-3em} G(\infty) = 0\,,\\
    &\omega^\prime(0) = 0\,, && \hspace*{-3em} \omega(\infty) = 0\,.
\end{align}
We numerically integrate these differential equations under the boundary conditions to determine the field profiles of $F(r)$, $G(r)$, and $\omega(r)$.



\section{Results}\label{sec:results}
We first show the analytical expressions of the distributions of the energy density, the pressure, and the shear force using the canonical EMT and the Belinfante EMT forms.
Next, we present the numerical results of those distributions.

\subsection{Analytical expressions}
Using Eqs.~\eqref{eq:Tcan} and \eqref{eq:Lagrangian}, we derive the canonical EMT in the present Skyrme model.

\begin{widetext}
The energy density is
\begin{equation}
    \begin{split}
    \varepsilon_\text{can}(r) &= \frac{1}{2}f_\pi^2\left({F^\prime}^2 + 2\frac{\sin^2 F}{r^2}\right) + f_\pi^2m_\pi^2(1 - \cos F) + \frac{2}{r^2}f_\pi^2(G + 1 - \cos F)^2 - f_\pi^2g^2\omega^2 \\ 
    &\qquad + \frac{1}{2g^2r^4}\left[2r^2{G^\prime}^2 + G^2(G + 2)^2\right] - \frac{1}{2} {\omega^\prime}^2 + \frac{3g\omega}{4\pi^2r^2}F^\prime\sin^2 F\,,
    \end{split}
\end{equation}
the pressure is
\begin{equation}
    \begin{split}
    p_\text{can}(r) &= - \frac{1}{6}f_\pi^2\left({F^\prime}^2 + 2\frac{\sin^2 F}{r^2}\right) - f_\pi^2m_\pi^2(1 - \cos F) - \frac{2}{3r^2}f_\pi^2(1 - \cos F)(G + 1 - \cos F) \\
    &\qquad - \frac{2}{r^2}f_\pi^2G(G + 1 - \cos F)+ f_\pi^2g^2\omega^2 - \frac{1}{6g^2r^4}\Bigl[2r^2{G^\prime}^2 + 3G^3(G + 2) + 2G^2(G + 2)\Bigr] + \frac{1}{6}{\omega^\prime}^2\,,
    \end{split}
\end{equation}
and the shear force is
\begin{equation}
    \begin{split}
    s_\text{can}(r) &= f_\pi^2\left({F^\prime}^2 - \frac{\sin^2 F}{r^2}\right) - \frac{2}{r^2}f_\pi^2(1 - \cos F)(G + 1 - \cos F) + \frac{1}{g^2r^4}\Bigl[2r^2{G^\prime}^2 - G^2(G + 2) - 3rGG^\prime\Bigr] - {\omega^\prime}^2\,.
    \end{split}
\end{equation}
The Belinfante versions of the energy density, the pressure, and the shear force read
\begin{equation}
    \begin{split}
    \varepsilon_\text{Bel}(r) &= \frac{1}{2}f_\pi^2\left({F^\prime}^2 + 2\frac{\sin^2 F}{r^2}\right) + f_\pi^2m_\pi^2(1 - \cos F) + \frac{2}{r^2}f_\pi^2(G + 1 - \cos F)^2 + f_\pi^2g^2\omega^2 \\
    &\qquad + \frac{1}{2g^2r^4}\left[2r^2{G^\prime}^2 + G^2(G + 2)^2\right] + \frac{1}{2} {\omega^\prime}^2\,,
    \end{split}
\end{equation}
\begin{equation}
    \begin{split}
    p_\text{Bel}(r) &= -\frac{1}{6}f_\pi^2\left({F^\prime}^2 + 2\frac{\sin^2 F}{r^2}\right) - f_\pi^2m_\pi^2(1 - \cos F) - \frac{2}{3r^2}f_\pi^2(G + 1 - \cos F)^2 + f_\pi^2g^2\omega^2 \\
    &\qquad + \frac{1}{6g^2r^4}\left[2r^2{G^\prime}^2 + G^2(G + 2)^2\right] + \frac{1}{6}{\omega^\prime}^2\,,
    \end{split}
\end{equation}
\begin{equation}
    \begin{split}
    s_\text{Bel}(r) &= f_\pi^2\left({F^\prime}^2 - \frac{\sin^2 F}{r^2}\right) - \frac{2}{r^2}f_\pi^2(G + 1 - \cos F)^2 + \frac{1}{g^2r^4}\left[r^2{G^\prime}^2 - G^2(G + 2)^2\right] - {\omega^\prime}^2\,.
    \end{split}
\end{equation}
\end{widetext}
In previous studies~\cite{Jung:2013bya, Fukushima:2020cmk}, these quantities were discussed, but we find inconsistent conventions.
The energy density was chosen to be the canonical form, while the pressure and the shear force were the Belinfante form.

The difference between the two EMT forms is attributed to the pseudo-gauge transformation with the spin current operator, $S^{\lambda\mu\nu}$.
In this model, its explicit form is
\begin{equation}
    S^{\lambda\mu\nu} = - \frac{2}{g^2}\tr(F^{\lambda\mu}V^\nu - F^{\lambda\nu}V^\mu)\,.
    \label{eq:spin}
\end{equation}
It should be noted that the spatial spin components, $S^{0ij}$, vanish because $F^{0i}$ involves only the $\omega$ meson, while $V^j$ represents the $\rho$ meson due to Eq.~\eqref{eq:omega}.

\subsection{Numerical results}
In our treatment of the Skyrme model, there are three fitting parameters, i.e., the pion decay constant, $f_\pi$, the pion mass, $m_\pi$, and the coupling constant, $g$.
We set $f_\pi$ and $m_\pi$ to the physical values:
\begin{equation}
    f_\pi=\SI{92.2}{MeV}\,,\qquad m_\pi=\SI{138}{MeV}\,.
    \label{eq:fpi}
\end{equation}
We fix $g=6.0$ to satisfy the KSFR (Kawarabayashi-Suzuki~\cite{Kawarabayashi:1966kd} and Fayyazuddin-Riazuddin~\cite{Riazuddin:1966sw}) relation, i.e., $2g^2f_\pi^2 = m_\rho^2 \simeq m_\omega^2 \simeq (\SI{783}{MeV})^2$.

\begin{figure}
    \centering
    \includegraphics[width=\linewidth]{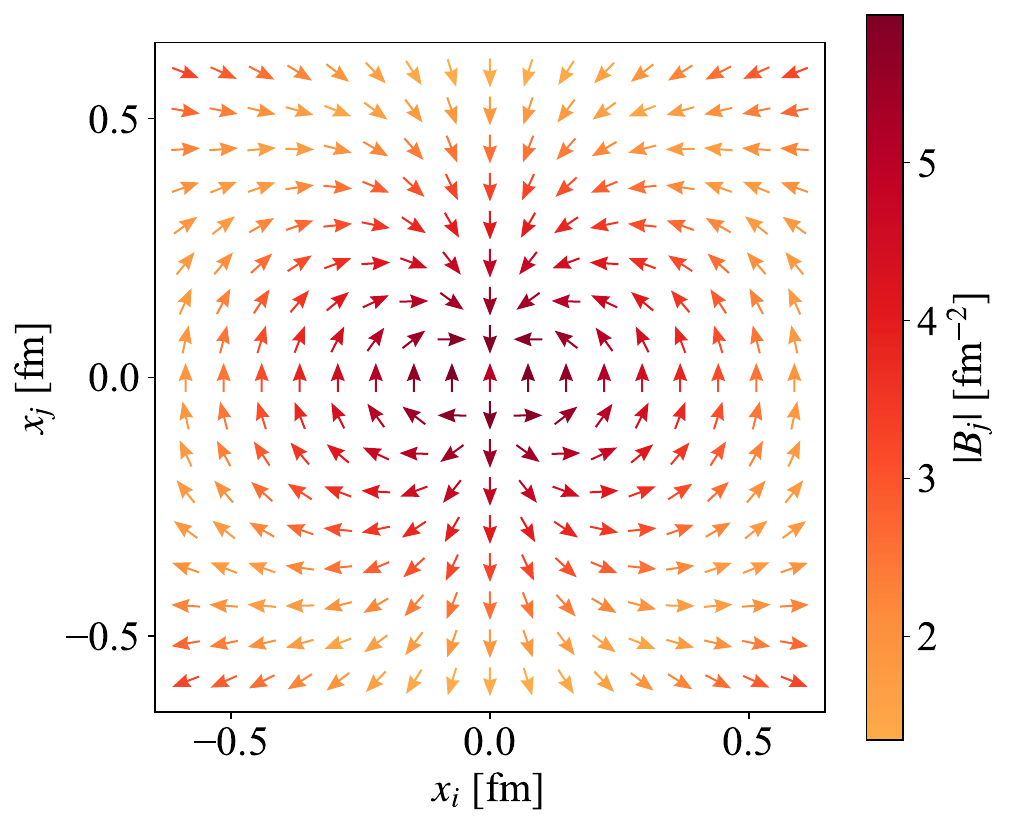}
    \caption{Distributions of the vector-meson magnetic profile in the flavor-$j$ component.  The $\rho$-meson Ansatz leads to the axial symmetric profile with respect to the $x_j$ axis.}
    \label{fig:magnetic}
\end{figure}

\begin{figure}
    \centering
    \includegraphics[width=\linewidth]{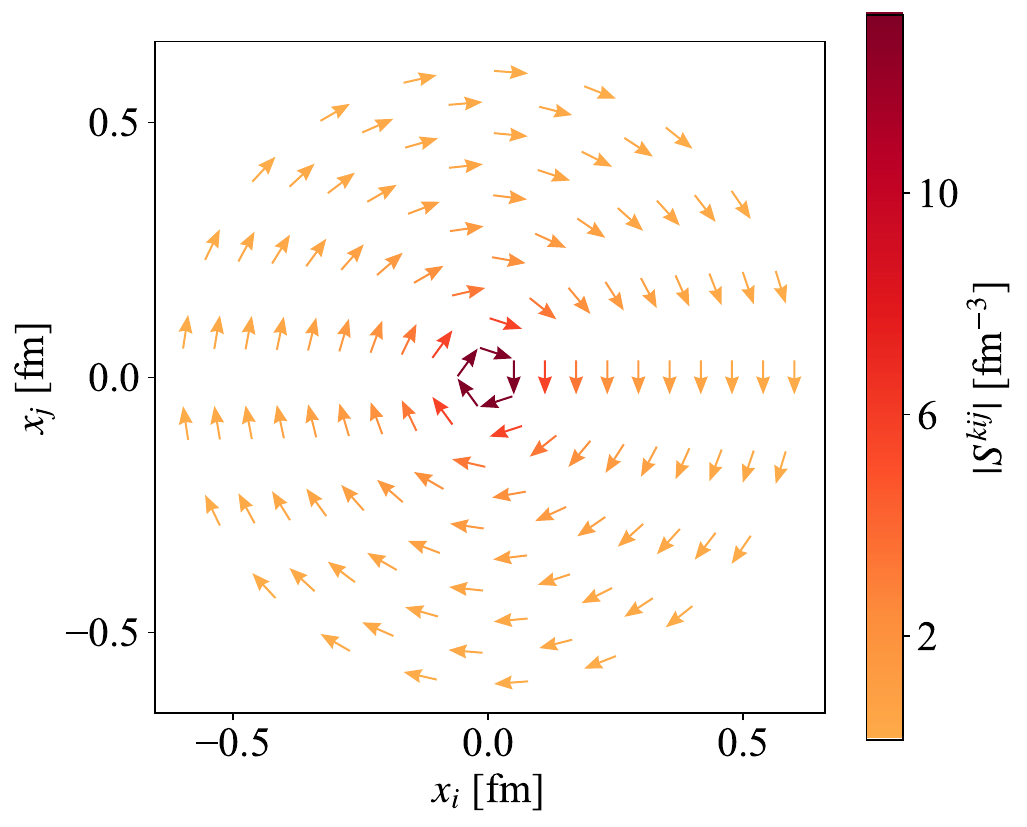}
    \caption{Distributions of the spin current in the $k$ direction. The corresponding spin charge density is $S^{0ij}$, which vanishes in our vector-meson Ans\"atze.}
    \label{fig:spin}
\end{figure}

First of all, let us begin with the spin current expectation value in Eq.~\eqref{eq:spin}.  We see that the vector-meson field strength tensors, $F^{\mu\nu}$, would induce a nonzero spin current, and it is interesting to consider the concrete profile of $F^{\mu\nu}$ surrounding the nucleon as shown in Fig.~\ref{fig:magnetic}.  We decompose the flavor structure on the basis of the Pauli matrices, and $B_j$ in Fig.~\ref{fig:magnetic} represents the magnetic distribution in the flavor-$j$ component, which is axial symmetric with respect to the $x_j$ axis.  This peculiar distribution is a consequence of the Ansatz~\eqref{eq:ansatz}.  Then, we present the spin current driven by the magnetic fields in Fig.~\ref{fig:spin}.  The plotted $S^{kij}$ in Fig.~\ref{fig:spin} represents the current of the spin charge density, $S^{0ij}$, flowing along $k$ direction.  This flow has no component perpendicular to the $x_i$-$x_j$ plane in our model setup.  Although the nucleon is spherically symmetric, the plot appears to show a preferred sense of rotation in Fig.~\ref{fig:spin}.  This originates from Ansatz~\eqref{eq:ansatz}: to stabilize the soliton, the $\rho$-meson field acquires a spatial profile with a definite rotational orientation.

\begin{figure}
    \centering
    \includegraphics[width=\linewidth]{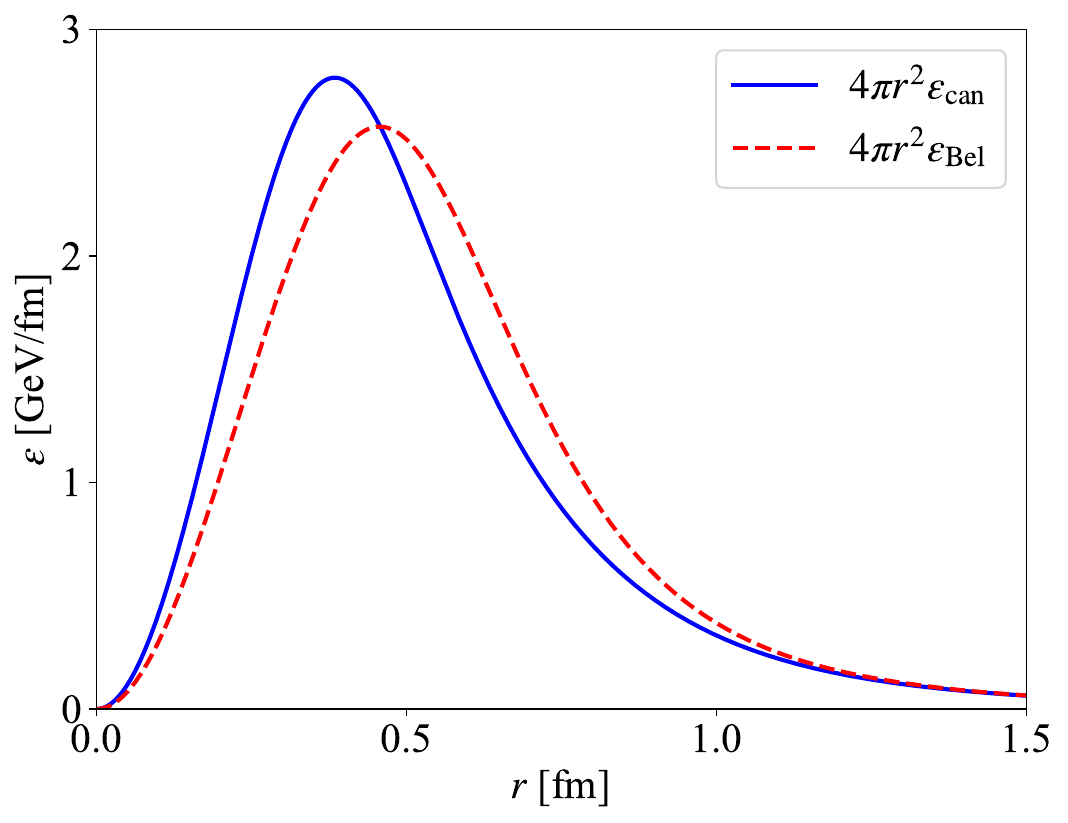}
    \caption{Distributions of the energy density in two different forms multiplied by $4\pi r^2$ as functions of $r$.}
    \label{fig:energy}
\end{figure}

We show the numerical results of the energy density in Fig.~\ref{fig:energy}.
We can see that the local values of the energy density are different for $\varepsilon_\text{can}$ and $\varepsilon_\text{Bel}$, while the total energy, 
\begin{equation}\label{eq:mass}
    \int_0^\infty\dd{r}4\pi r^2\varepsilon(r) = \SI{1.45}{GeV}\,,
\end{equation}
is insensitive to the pseudo-gauge choice, which is understood from the explicit form:
$\varepsilon_\text{Bel} - \varepsilon_\text{can} = -(2/g^2)\partial_\lambda\tr(F^{\lambda0}V^0)$.  We note that this total energy is identified as the nucleon mass, though it overshoots the physical value $\sim \SI{0.94}{GeV}$ of the proton and neutron mass.  In the model, we could reproduce the nucleon mass, but then $f_\pi$ turns out to be unphysically small~\cite{Adkins:1983nw}.  In later analyses, we will introduce a prescription to rescale the mass to the physical value.

In view of Fig.~\ref{fig:energy}, $\varepsilon_\text{can}$ appears to be more localized near the center than $\varepsilon_\text{Bel}$.  One may thus think that the nucleon size would depend on the pseudo-gauge choice.  However, it is physically unnatural that the pseudo-gauge choice would affect any measurable quantities.  In fact, we remark that the nucleon size can be characterized by the radii based on the distribution of either the baryon density (baryon density radius) or the vector mesons (isoscalar/isovector radius), which are pseudo-gauge independent as they should.

\begin{figure}
    \centering
    \includegraphics[width=\linewidth]{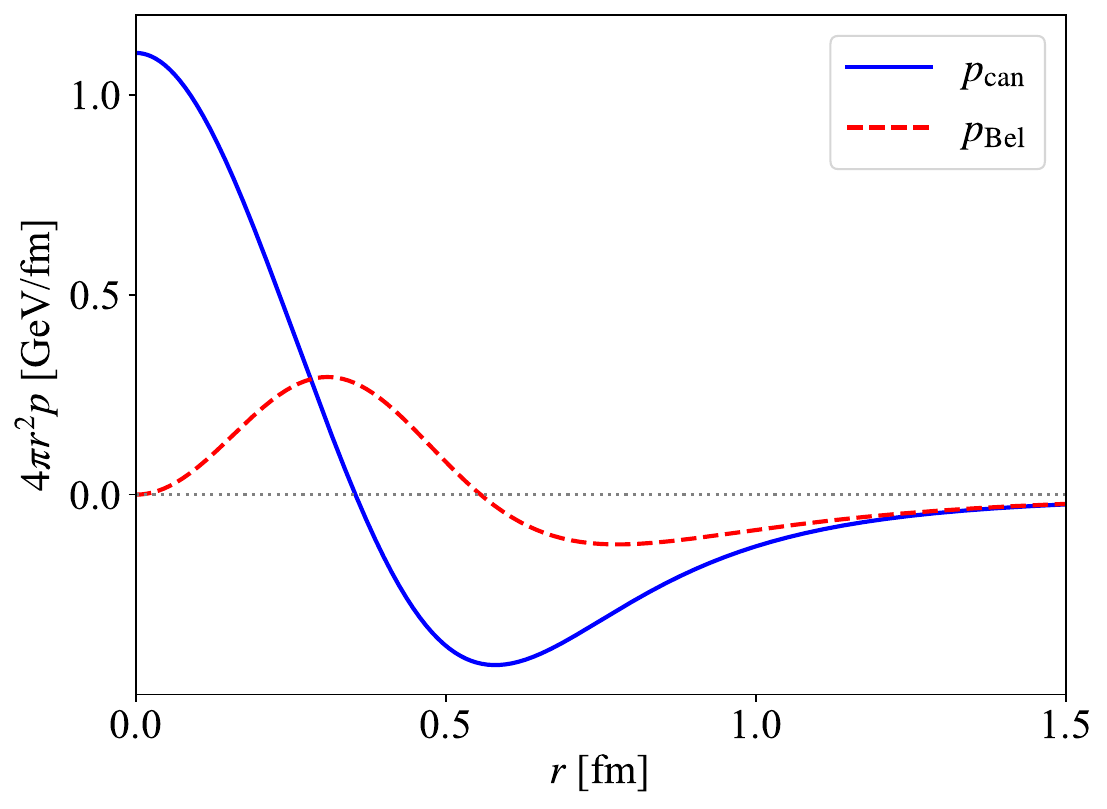}
    \caption{Distributions of the pressure in two different forms multiplied by $4\pi r^2$ as functions of $r$.}
    \label{fig:pressure}
\end{figure}

We next proceed to the numerical results of the pressure in Fig.~\ref{fig:pressure}.  For the pressure the von~Laue condition must hold~\cite{Polyakov:2018zvc, Lorce:2018egm}, which guarantees pseudo-gauge independence of the integrated quantities; that is,
\begin{equation}\label{eq:vonLaue}
    \int_0^\infty\dd{r}r^2p_\text{can}(r) = \int_0^\infty\dd{r}r^2p_\text{Bel}(r) = 0\,.
\end{equation}
Although the difference in the explicit expressions takes the form of the surface term like the energy density, the pressure exhibits a drastic change at the qualitative level.

Figure~\ref{fig:pressure} shows the distributions of the different pressures, $p_\text{can}$ and $p_\text{Bel}$, as functions of $r$.  We clearly notice from a comparison between Figs.~\ref{fig:energy} and \ref{fig:pressure} that the pressure has much larger pseudo-gauge dependence.  In particular, the limiting behavior at $r\to 0$ qualitatively changes as
\begin{equation}
    p_\text{can}(r\to 0) \sim r^{-2}\,,\qquad
    p_\text{Bel}(r\to 0) \sim r^0\,.
    \label{eq:przero}
\end{equation}
The singular behavior in $p_\text{can}(r)$ near the center causes nontrivial dependence of the integrated quantities on the pseudo-gauge choice.  To see this, we shall generalize the von~Laue condition~\eqref{eq:vonLaue}~\cite{Lorce:2018egm}.  For any sufficiently smooth function, $f(r)$, we easily prove the following relation from Eq.~\eqref{eq:hydroequilibrium}:
\begin{align}
     f(r)p_r(r)\Bigr|_{r=0}^{r=\infty} 
     &= \int_0^\infty\dd{r} \frac{\dd}{\dd r} \left[f(r) p_r(r) \right] \notag\\
     &= \int_0^\infty\dd{r} \left[ f'(r) p_r(r) - 2\frac{f(r)}{r}s(r)\right]\,.
\end{align}
The left-hand side vanishes for $f(0)p_r(0)=0$ and $f(r)p_r(r)|_{r\to \infty}\to 0$.  Now, we choose $f(r) = r^N$ for some real number $N$ and require that $p_r(r)$ drops sufficiently fast, we eventually get
\begin{align}
    0 &= \int_0^\infty\dd{r} r^{N-1}\left[Np_r(r) - 2s(r)\right] \notag\\
    &= \int_0^\infty\dd{r} r^{N-1} \left[ (N-2)p_r(r) + 2p_{\theta,\phi}(r) \right]
\end{align}
thanks to $p_r(r)-p_{\theta,\phi}(r)=s(r)$.

For $N=3$ we can immediately retrieve the von~Laue condition~\eqref{eq:vonLaue}.  We can arbitrarily increase $N$ as long as the $r$ integration converges; for example, the sum rule for $N=4$ reads
\begin{equation}
    \int_0^\infty \dd{r}\, r^3\, \left[ p_r(r) + p_{\theta,\phi}(r) \right] = 0
\end{equation}
and the choice of $N=2$ leads to
\begin{equation}
    \int_0^\infty \dd{r}\, r\, p_{\theta,\phi,\text{Bel}}(r) = 0\,.
\end{equation}
It should be noted that the latter expression for $N=2$ makes sense only for the Belinfante form; otherwise, the canonical pressure is singular at $r\to 0$ as pointed out in Eq.~\eqref{eq:przero}.
For the same reason, we cannot discuss the $N=1$ relation for the canonical form, while the Belinfante form still works.

\begin{figure}
    \centering
    \includegraphics[width=\linewidth]{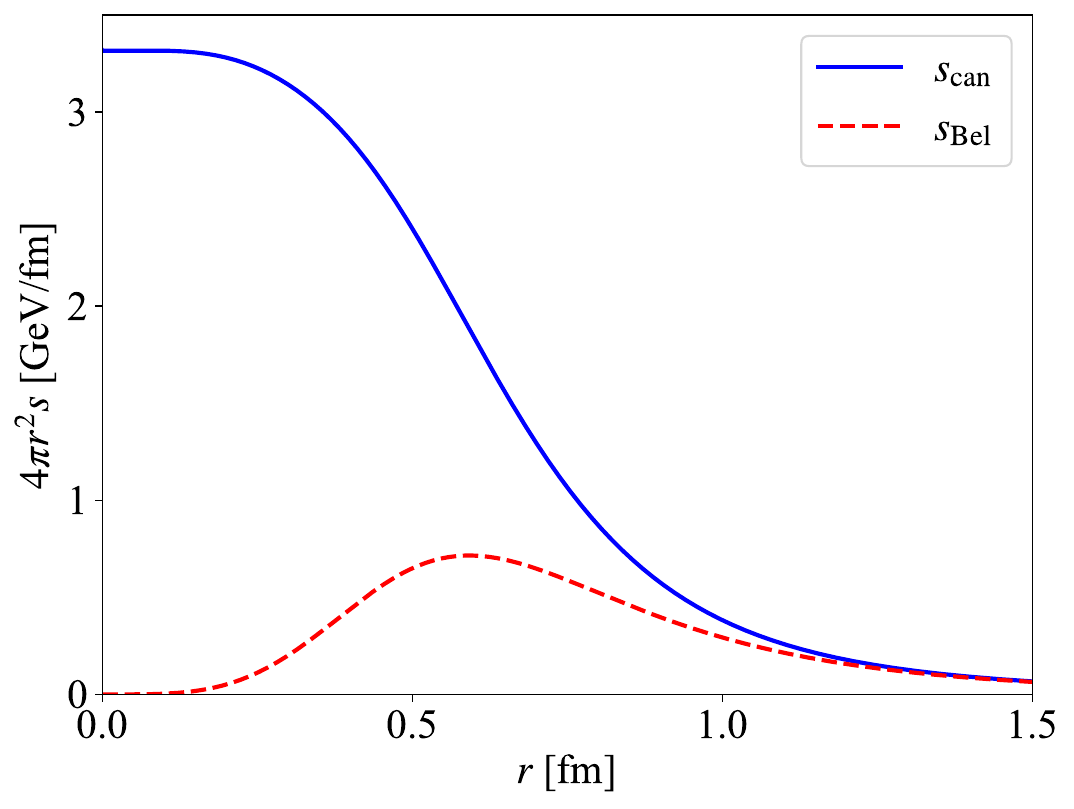}
    \caption{Distributions of the shear force in two different forms multiplied by $4\pi r^2$ as functions of $r$.}
    \label{fig:shear}
\end{figure}

Finally, we discuss the numerical results of the shear force in Fig.~\ref{fig:shear}.  In contrast to the energy density and the pressure, the difference in the shear force does not take the form of the surface term.  The volume integration of the energy-momentum tensor is pseudo-gauge independent simply because of the tensorial structure of the terms involving $s(r)$ in Eq.~\eqref{eq:EMT}.  That is,
\begin{equation}
    \int \dd[3]{\bm{r}} \left(\frac{r^ir^j}{r^2} - \frac{1}{3}\delta^{ij}\right)s(r) = 0\,,
\end{equation}
for any $s(r)$.  Yet, it may be meaningful to define the radial integration of the shear force, which is sometimes called the ``surface tension energy'' of the nucleon~\cite{Polyakov:2018zvc}.  Then, the surface tension energy depends on the pseudo-gauge choice; in our numerical results, we find:
\begin{align}
    &\int_0^\infty\dd{r} 4\pi r^2 s_\text{can}(r) = \SI{2.3}{GeV}\,,\\
    &\int_0^\infty\dd{r} 4\pi r^2 s_\text{Bel}(r) = \SI{0.48}{GeV}\,.
\end{align}
At a glance of Fig.~\ref{fig:shear}, the area of $4\pi r^2\,s_\text{can}(r)$ is far wider than that of $4\pi r^2\,s_\text{Bel}$, as confirmed by the above numerical values.
We note that $s_\text{can}(r\to 0)\sim r^{-2}$ is inevitably concluded from $\varepsilon_\text{can}(r\to 0) = \text{(const.)}$ and $p_\text{can}(r\to 0)\sim r^{-2}$ in order to satisfy Eq.~\eqref{eq:hydroequilibrium}.

\section{Discussions}\label{sec:discussions}

We shall exemplify the problems of pseudo-gauge ambiguity by considering physical quantities as often discussed in the previous studies.  First, we compute the confining force inside the nucleon using the canonical and Belinfante forms.  Next, we estimate the EoS of single-nucleon matter at high energy density.  Both quantities suffer pseudo-gauge dependence.

\subsection{Confining properties}
\begin{figure}
    \centering
    \includegraphics[width=\linewidth]{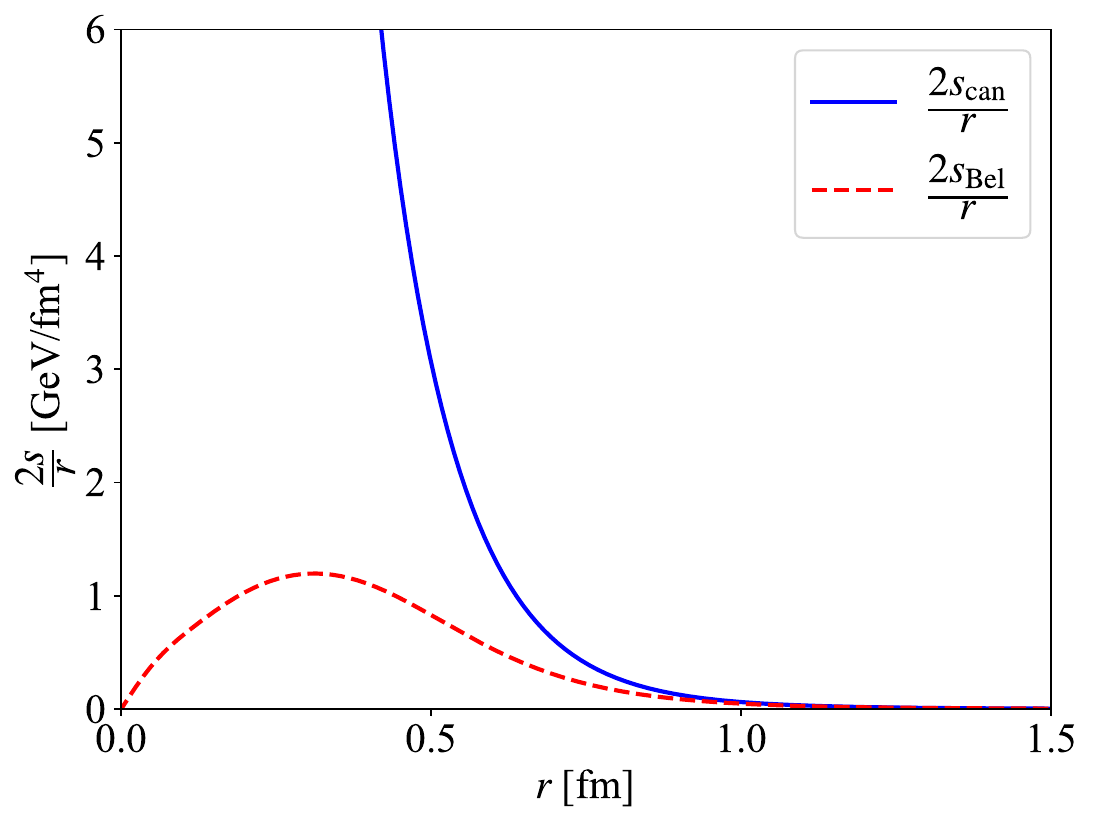}
    \caption{Distributions of the confining force in two different forms as functions of $r$.}
    \label{fig:force}
\end{figure}

From the hydroequilibrium condition, we can identify the confining force from the right-hand side of Eq.~\eqref{eq:hydroequilibrium}, which resists the pressure gradient, that is,
\begin{equation}
    f_{\text{confining}}(r) = -\frac{2}{r} s(r)\,.
    \label{eq:f_conf}
\end{equation}
Thus, if we experimentally access the distribution of the shear force inside the nucleon, it would be feasible to determine the strength of the confining force quantitatively.  Here, we should emphasize that the above argument is the robust way to define the confining force.  It is often said that the negative pressure as seen in Fig.~\ref{fig:pressure} should be interpreted as a confining force near the surface of the nucleon in accord with the bag model picture of confinement.  However, the negative pressure region inevitably arises to satisfy the von~Laue condition that is guaranteed merely by the virial theorem.  If we look at $p_\text{Bel}(r)$ in Fig.~\ref{fig:pressure}, the negative pressure region is extended to $r\gtrsim \SI{0.7}{fm}$, but the peak in the confining force in Fig.~\ref{fig:force} is located around $r\sim \SI{0.3}{fm}$.  Thus, the negative pressure is necessary for mechanical stability, but it may not be directly related to the confining properties, and the relevant length scales turn out to be separate.

Although the force in Eq.~\eqref{eq:f_conf} has a clear meaning, the confining force suffers substantial pseudo-gauge dependence.  We already pointed out that the length scales are found to differ in the Belinfante form, but the situation in the canonical form is even disastrous.  As noticed in Fig.~\ref{fig:force}, the confining force derived from the canonical EMT diverges at small $r$, and this divergent force is equated by a strongly enhanced pressure gradient.  Therefore, mathematically, consistency is not violated, but it is far from obvious whether the cancellation between huge contributions is physically sensible.  Our results suggest that the Belinfante EMT looks more reasonable.  Nevertheless, there is no \textit{a priori} criterion that could justify or falsify any particular form of the EMT\@.

\subsection{EoS of single-nucleon matter}
If $p(r)$ and $\varepsilon(r)$ are simultaneously determined, we can deduce the EoS, $p=p(\varepsilon)$, in matter inside the nucleon by eliminating the parameter $r$.
Since dense nuclear matter is saturated eventually by overlapping wave-functions of nucleons, it is conceivable to approximate the bulk properties of densely saturated matter by the nucleon properties.  We note that the nuclear interaction may be dominant at low energies, and the above approximation would work at high enough density where quark degrees of freedom are liberated.

It is quantitatively important to note that the nucleon mass is overestimated in the present soliton model.  According to the prescription in Ref.~\cite{Fukushima:2020cmk}, we rescale the energy density and the pressure density by a ratio parameter read from
\begin{equation}
    \chi = \frac{(\text{physical mass})}{(\text{model mass})} \approx \frac{\SI{940}{MeV}}{\SI{1450}{MeV}} \approx 0.65
\end{equation}
as
\begin{equation}
    \varepsilon(r) \to \chi\,\varepsilon(r)\,,
    \qquad
    p(r) \to \chi^{-1}\, p(r)\,.
\end{equation}
Since the nucleon mass is the integral of the energy density, the first scaling is natural.  The second scaling for $p(r)$ is fixed by the requirement that the EMT form factors remain unchanged~\cite{Fukushima:2020cmk}.

\begin{figure}
    \centering
    \includegraphics[width=\linewidth]{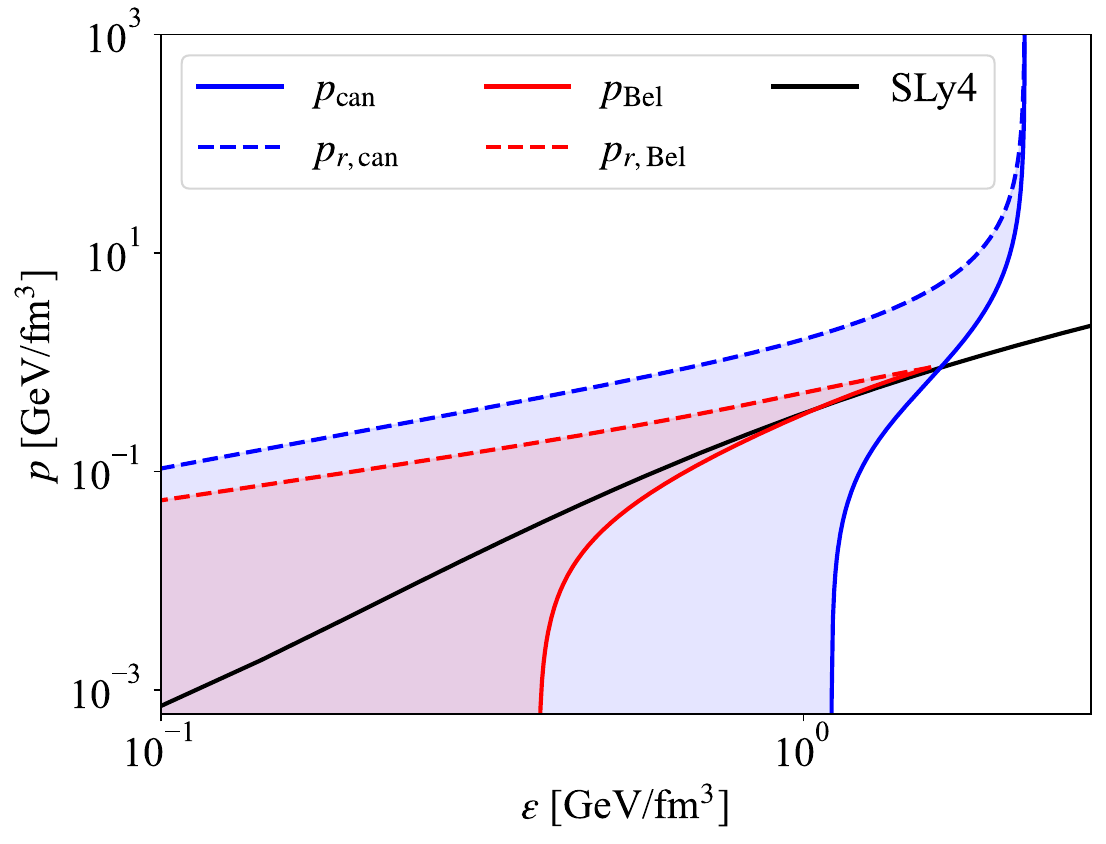}
    \caption{EoSs inside the single nucleon obtained in several different prescriptions.  The canonical (blue lines) and Belinfante (red lines) results show additional ambiguity in the pressure definition; $p$ or $p_r$.  The (black) solid line represents an empirical EoS of NS matter called the SLy4.}
    \label{fig:eos}
\end{figure}

Figure~\ref{fig:eos} summarizes our results from the EoS calculations.  As argued in Refs.~\cite{Rajan:2018zzy, Fukushima:2020cmk}, the single-nucleon EoS was found to be similar to the empirical EoS of neutron star (NS) matter.
Indeed, we have verified this observation;  Fig.~\ref{fig:eos} shows a typical example of the NS matter EoS, namely, SLy4, by the black solid line.  Remarkably, $p_\text{Bel}(\varepsilon)$ is pretty well matched with the SLy4 in the high density region.  It is reasonable that $p_\text{Bel}(\varepsilon)$ agrees better with NS matter EoS because the standard calculation scheme for the NS matter EoS is based on the Belinfante EMT~\cite{Chatterjee:2014qsa}.  Also, we note that a similar pseudo-gauge nonuniqueness of the EMT in the massive gravitational field theory is discussed in Ref.~\cite{Bicak:2016slm}.

It is, however, subtle whether the single-nucleon EoS can provide a baseline.  The pseudo-gauge dependence is not negligible as manifested in the comparison in Fig.~\ref{fig:eos}.  Even if we adopt the Belinfante form and discard the canonical one for some unknown reason, the EoS is not unique.
This non-uniqueness problem is caused by the shear force.  In the case of NS matter, the anisotropy in the pressure distribution is a minor effect, but in the nucleon interior, as illustrated in Fig.~\ref{fig:pr_ptheta}, the shear force makes a significant deviation between $p(r)$ and $p_r(r)$.
It is a nontrivial question which of $p(r)$ and $p_r(r)$ should correspond to the EoS that is relevant for quark matter properties.
Of course, in a finite-volume and strongly inhomogeneous system like a nucleon, there is no rigorous justification for such a prescription to approximate the bulk thermodynamic EoS by the local internal properties of the nucleon.

\section{Conclusions}\label{sec:conclusions}
We have investigated how pseudo-gauge ambiguity of the energy–momentum tensor (EMT) impacts local mechanical properties of the nucleon within a two-flavor Skyrme model with vector mesons. The canonical and Belinfante forms differ by total derivative terms, which are associated with the spin current constructed from the vector–meson field strengths.  In our setup, the spin charge $S^{0ij}$ vanishes, while the spin current $S^{kij}$ is driven by spatially inhomogeneous profiles of the $\rho$ and $\omega$ mesons, leading to distinct local distributions of conserved charge densities.

Consistently, the energy density distributions, $\varepsilon_\text{can}(r)$ and $\varepsilon_\text{Bel}(r)$, are moderately shifted by surface terms, so that the integrated energy, i.e., the soliton mass, is pseudo-gauge invariant, while the local profiles differ.  In contrast, the pressure is far more sensitive;  the small-$r$ behavior changes qualitatively from $p_\text{can}\sim r^{-2}$ to $p_\text{Bel}\sim r^{0}$, which in turn modifies families of pressure sum rules beyond the standard von~Laue relation.
The shear force tensor integrates to zero in the volume, independent of pseudo-gauge, but the commonly discussed ``surface tension energy'' given by $\int \dd{r} 4\pi r^2 s(r)$ is not invariant and differs substantially between the two EMT forms.

Guided by hydrostatic equilibrium, we identified a kinematic measure of the confining force, $f_\text{confining}(r) = -2s(r) / r$.  The peak location of the confining force is not tied to the radius where the pressure turns negative; the negative-pressure region required by mechanical stability may extend to larger $r$.  Thus, the interpretation of negative pressure may have theoretical subtlety beyond the bag-model picture.  Actually, in the canonical pseudo-gauge, $f_\text{confining}(r)$ diverges at small $r$ and is only compensated by a large pressure gradient, which could be a mathematically consistent but physically dubious cancellation.  The Belinfante EMT looks much more reasonable, but to our best knowledge, there is no \textit{a priori} principle that uniquely selects a pseudo-gauge.

Finally, after rescaling $\varepsilon$ and $p$ to account for the nucleon mass overestimated in the model, we examined a ``single-nucleon'' equation of state (EoS).  The Belinfante-based EoS aligns with a representative NS EoS (SLy4) at high energy density, but a non-negligible pseudo-gauge dependence persists.  More importantly, strong shear in the nucleon interior makes the isotropic pressure $p(r)$ and the radial pressure $p_r(r)$ inequivalent, so even within one fixed pseudo-gauge, the inferred EoS is not unique.

Taken together, our results demonstrate that while global constraints such as the mass, the von~Laue condition, etc.\ are robust, local mechanical properties, i.e., energy density, pressure, shear, surface tension energy, confining force, etc.\ are pseudo-gauge sensitive.  This issue may have been overlooked or underestimated because the pseudo-gauge dependence gets pronounced only when vector fields are involved in the model.  Thus, it was essential for us to study this issue by means of the vector-meson-extended Skyrme model.  This calls for caution when mapping GPD/DVCS information onto local structures of the nucleon.
One might think that, in view of our present analyses, the Belinfante EMT is preferable and singular results from the canonical EMT are physically irrelevant.  However, one may encounter exceptions if not only the forward limit is concerned.  Indeed, we know that the canonical EMT is convenient for investigating the spin contents, cf., the Jaffe-Manohar spin decomposition is based on the canonical EMT\@.  Interesting questions, such as identifying pseudo-gauge–invariant characterizations~\cite{Becattini:2025twu} and seeking theoretical or phenomenological principles of selecting the preferable EMT form, deserve further investigations in the future.

\begin{acknowledgments}
This work was partially supported by Japan Society for the Promotion of Science
(JSPS) KAKENHI Grant No.\ 
22H01216 (K.F.) and
FoPM, WINGS Program, the University of Tokyo (T.U.).
\end{acknowledgments}

\bibliography{shear}
\end{document}